# A big-world network in ASD: Dynamical connectivity analysis reflects a deficit in long-range connections and an excess of short-range connections.

Pablo Barttfeld[1,2], Bruno Wicker [1,3], Sebastián Cukier[1,2], Silvana Navarta[1] , Sergio Lew[4] and Mariano Sigman[1]

[1]Integrative Neuroscience Laboratory, Physics Department, University of Buenos Aires.

[2]Fundación para la Lucha contra las Enfermedades Neurológicas de la Infancia, Buenos Aires.

[3]Mediterranean Institute of Cognitive Neurosciences, Centre National de la Recherche Scientifique, Université de la Méditerranée, 31 Chemin Joseph Aiguier, 13402 Marseille Cedex 20, France.

[4]Instituto de Ingeniería Biomédica, Facultad de Ingeniería, Universidad of Buenos Aires

Corresponding author: Pablo Barttfeld

Integrative Neuroscience Laboratory, Physics Department, University of Buenos Aires, Buenos Aires, Argentina.

e-mail:pbarttfeld@fi.uba.ar

**Abstract**

Over the last years, increasing evidence has fuelled the hypothesis that Autism Spectrum Disorder (ASD) is a condition of altered brain functional connectivity. The great majority of these empirical studies rely on functional magnetic resonance imaging (fMRI) which has a relatively poor temporal resolution. Only a handful of studies have examined networks emerging from dynamic coherence at the millisecond resolution and there are no investigations of coherence at the lowest frequencies in the power spectrum – which has recently been shown to reflect long-range cortico-cortical connections. Here we used electroencephalography (EEG) to assess dynamic brain connectivity in ASD focusing in the low-frequency (delta) range. We found that connectivity patterns were distinct in ASD and control populations and reflected a double dissociation: ASD subjects lacked long-range connections, with a most prominent deficit in fronto-occipital connections. Conversely, individuals with ASD showed increased short-range connections in lateral-frontal electrodes. This effect between categories showed a consistent parametric dependency: as ASD severity increased, short-range coherence was more pronounced and long-range coherence decreased. Theoretical arguments have been proposed arguing that distinct patterns of connectivity may result in networks with different efficiency in transmission of information. We show that the networks in ASD subjects have less Clustering coefficient, greater Characteristic Path Length than controls -indicating that the topology of the network departs from small-world behaviour- and greater modularity. Together these results show that delta-band coherence reveal qualitative and quantitative aspects associated with ASD pathology.

# 1. Introduction

Autism or Autism Spectrum Disorder (ASD) is a neurodevelopmental disorder characterized by a triad of impairments in social interaction, communication, and behavioural flexibility (APA, 2000). There is increasing evidence that ASD could be a condition of altered brain connectivity (Belmonte et al., 2004; Courchesne & Pierce, 2005; Just, Cherkassky, Keller, Kana, & Minshew, 2007; Just, Cherkassky, Keller, & Minshew, 2004; Markram, Rinaldi, & Markram, 2007; Wicker, 2008). Anatomical studies showed that individuals with ASD have smaller and more densely packed columns of neuronal cells (Casanova & Trippe, 2009; Casanova et al., 2006; Hughes, 2007). Structural MRI studies have reported a reduced corpus callosum (Alexander et al., 2007; Egaas, Courchesne, & Saitoh, 1995) and abnormal anatomy and connections of the limbic–striatal social brain system in ASD (McAlonan et al., 2005). fMRI studies also yielded evidence of altered connectivity: connectivity within the frontal lobe seems differently organized, and areas such as prefrontal cortex, precuneus/posterior cingulate cortex and superior temporal sulcus, appear to be poorly connected (Just, Cherkassky, Keller, Kana, & Minshew, 2007; Just, Cherkassky, Keller, & Minshew, 2004; Kana, 2006; Kleinhans et al., 2008; Koshino et al., 2005; Mason, Williams, Kana, Minshew, & Just, 2008; Welchew et al., 2005; Wicker, 2008). Recent fMRI studies have moved away from the social and cognitive deficit models and looked at the functional connectivity between areas of the so-called default mode network (DMN), i.e. networks that become activated at rest (Gusnard & Raichle, 2001; Raichle, 2009) Results revealed decreased connectivity between the medial prefrontal cortex and precuneus/posterior cingulate cortex (Cherkassky, 2006; Di Martino et al., 2009; Weng, 2010). The great majority of the evidence signalling functional connectivity as a key aspect of ASD was obtained using fMRI. There are only a handful of studies assessing ASD brain connectivity using the other canonical tool to study connectivity, electroencephalography (EEG). Abnormal gamma activity has been reported in autistic children, interpreted as supporting hypotheses of abnormal connectivity (Brown, Gruber, Boucher, Rippon, & Brock, 2005). Murias et al. (2007) and Coben et al. (2008) measured connectivity more directly using EEG coherence and reported evidence of both under- and over-connectivities in different frequency bands in ASD populations.

Understanding the patterns of connectivity in low frequencies - the delta band - remains unexplored.

Over the last years, analysis of coherence at low-frequencies has gained great interest. Long-range intra-cortical and feedback cortico-cortical connections, specifically those connections thought to be altered in ASD, contribute directly and significantly to the slow cortical potentials of the EEG (He & Raichle, 2009; He, Snyder, Zempel, Smyth, & Raichle, 2008). Also, the power coherence in low-frequency bands, particularly in the delta band correlates with the resting-state fMRI signal. (He, Snyder, Zempel, Smyth, & Raichle, 2008; Lu, 2007). Together, these results suggest that connectivity in the lower bands might specifically reflect the alterations of long range connectivity thought to play a key role in ASD.

Based on this hypothesis, we investigated whether functional brain networks in EEG are abnormally organized in the delta band in ASD. We will show that control subjects have stronger long fronto-occipital connections, and weaker lateral frontal connections and these differences are good predictors of ASD severity. Interestingly, the specificity of the connections is modulated by ASD severity (and not simply the amount of connections): short connections correlate positively with ASD (stronger connections relating to more severe ASD) whereas long connections correlate negatively with ASD (weaker connections relating to more severe ASD). When inspecting the impact of this connectivity pattern on the global organization of functional network using graph theory measures we observed that ASD present longer Characteristic Path Length (L), smaller Clustering Coefficient (C) and higher modularity index (MI), resulting in a less efficient brain network (Latora & Marchiori, 2001).

**2. Material and Methods.**

*2.1 Participants*

Two groups took part in this study. The ASD group included 10 adults with high-functioning autism or Asperger's syndrome (9 men and 1 women; mean age = 23.8, std = 7.6, Table 1). The individuals with ASD were provisionally accepted into the study if they had received a diagnosis of infantile autism or Asperger's syndrome from a child psychiatrist, developmental pediatrician, or licensed clinical psychologist. Actual participation required that this diagnosis had been recently confirmed, with each subject having met the criteria for ASD within the past 3 years on the basis of the revised fourth edition of the Diagnostic and Statistical Manual of Mental Condition (APA, 2000) and on the score on the Autistic Diagnostic Observation Schedule-Generic (Lord et al., 2000). IQs were measured with the third edition of the Wechsler Adult Intelligence Scale and ranged from 85 to 125 (mean=101.7, SD=14.97). At the time of testing, no ASD subject had known associated medical condition. As their mean IQ score was within the normal range, the ASD participants were individually matched to a group of 10 typically developing individuals on the basis of sex and chronological age. The participants in the control group (9 men, 1 women, mean age = 25.3, std = 6.54) were all volunteers and were free of psychiatric condition at the time of testing.

| Age | Sex | Diagnosis | Verbal IQ | Exec IQ | Total IQ | *ADOS* | | |
|---|---|---|---|---|---|---|---|---|
| | | | | | | Comm. | Soc. Int. | Total |
| 32 | F | Asperger | 116 | 78 | 99 | 3 | 4 | 7 |
| 26 | M | HFA | 98 | 80 | 90 | 10 | 5 | 15 |
| 22 | M | HFA | 96 | 74 | 85 | 3 | 9 | 12 |
| 17 | M | HFA | 111 | 127 | 120 | 5 | 8 | 13 |
| 17 | M | Asperger | 88 | 83 | 85 | 2 | 7 | 9 |
| 30 | M | HFA | 114 | 130 | 125 | 4 | 8 | 12 |
| 38 | M | HFA | 111 | 131 | 121 | 6 | 9 | 15 |
| 24 | M | HFA | 104 | 89 | 98 | 5 | 6 | 11 |
| 16 | M | HFA | 99 | 89 | 95 | 3 | 9 | 12 |
| 16 | M | HFA | 94 | 105 | 99 | 6 | 10 | 16 |

**Table 1**. Details of ASD subjects: age, diagnosis, IQ and ADOS scores.

*2.2 EEG recordings.*

EEG were recorded with a Biosemi Active-two 128 channels equipment, at a rate of 512 Hz. 7-minute temporal signals were recorded during an eyes-closed resting while subjects sat on a reclining chair in a sound attenuated room with a dim light. Digital, zero-phase shift filtering of the EEG temporal signals in the delta band (0.5-3.5Hz) was done performed offline in order to compute the correltion matrices. Correlations between all pair wise combinations of EEG channels were computed for all subjects with the Synchronisation Likelihood (SL) method (Stam & van Dijk, 2002). All parameters were set as described in detail in (Montez, Linkenkaer-Hansen, van Dijk, &

Stam, 2006). This resulted in a 128x128 symmetrical matrix for each subject, where each entry in the matrix represents the SL between the corresponding pair of electrodes. All subsequent analysis and statistics were performed from these SL matrices.

*2.3 Graph Theory metrics.*

The connectivity matrix defines a weighted graph where each electrode corresponds to a node and the weight of each link is determined by the SL of the electrode pair. To calculate network measures, SL matrices were converted to binary undirected matrices by applying a threshold $T$. We explored a broad range of values of $0.01 < T < 0.2$, with increments of 0.0005 and repeated the full analysis for each value of $T$. After transforming the SL-matrix to a binary undirected graph, we measured the Clustering Coefficient $C$, the Characteristic Path Length $L$ and Modularity Index $MI$ using the BCT toolbox (Rubinov & Sporns, 2009). For statistical comparisons of graph based metrics, we performed ANOVAs with group (control or ASD) and threshold (binned in 8) as independent factors. Errors were calculated using bootstrap (Efron & Tibshirani, 1994), which were used to explore statistical differences for all individual thresholds. The bootstrap probability was also calculated, by resampling 2000 times each metric for both groups, and calculating the percentage of times the mean metric of a group was larger than the mean metric of the other group. Network visualizations were performed using the Pajek software package (Batagelj & Mrvar, 1998) using a Kamada-Kawai layout algorithm (Kamada & Kawai, 1989).

**3. Results**

For each participant in this study, we calculated the Synchronization Likelihood (SL) across all pairs of channels (see methods and Montez et al., 2006 for details). SL provides a measure of temporal coherence between two temporal signals. This measure is more sensitive than simply a linear-correlation because: 1) it does not assume linearity in the coherence and 2) it is sensitive to phase-shifted coherent frequency bands which may result in a null linear correlation. This analysis collapsed the stationary EEG data of each participant, band passed in the delta range to a 128x128 synchronization matrix (henceforth referred as SL-matrix). The element *(i,j)* of the matrix provides a measure of the temporal similarity at low frequencies of electrodes *i* and *j* during eyes-closed stationary EEG, which we refer as functional connectivity. In what follows, we analyse statistical differences in functional connectivity for ASD and control population.

To calculate significant differences in SL patterns across groups, we conducted a paired t-test with the SL value for each pair of channels (Figure 1b). A positive t-value indicates that SL increased in control compared to ASD population. Conversely, a negative t-value indicates that SL is greater in ASD than in the control population. The distribution of t-values (Figure 1d) was significantly shifted towards positive values (mean= 0.57; std= 1,11; t-test: p=0, CImin= 0.55; CImax=0.58) indicating that the global trend was that SL was greater in the control population. Our interest was to understand the topography of the tails of this distribution, i.e. which pairs of electrodes had a greater difference in SL between ASD and control population. For this, we simply determined an arbitrary cut-off at $t = 2$ (Figures 1b and 1d) and considered the resulting matrix with values 1, 0 or -1 depending on whether $t > 2$, $2 > t > -2$ or $-2 > t$ (Figure 1f). This cut-off is certainly arbitrary but none of the results discussed in what follows depend on this choice (see Supplementary Figure 1 for a progression of the correlation topographies for varying thresholds). To further constrain the number of comparisons and generate a relatively sparse pattern of connections amenable to visualize its topography, we applied a mask, considering only pairs of electrodes with SL values exceeding a threshold of 0.03 (Figures 1c and 1e). The topographic projections of connections whose strength increased (red) or decreased (blue) in ASD compared to control participants (Figure 1h) showed a very consistent pattern. Connections which were stronger in the control group were localized in the frontal lobe and extended over the midline to the occipital cortex. They also included long-range connections connecting these regions. On the contrary, connections which were stronger in the ASD group were very focal and largely localized to the lateral frontal electrodes. These observations did not change qualitatively when changing the thresholds of the binary difference matrix or the activation mask (Supplementary Figure 1).

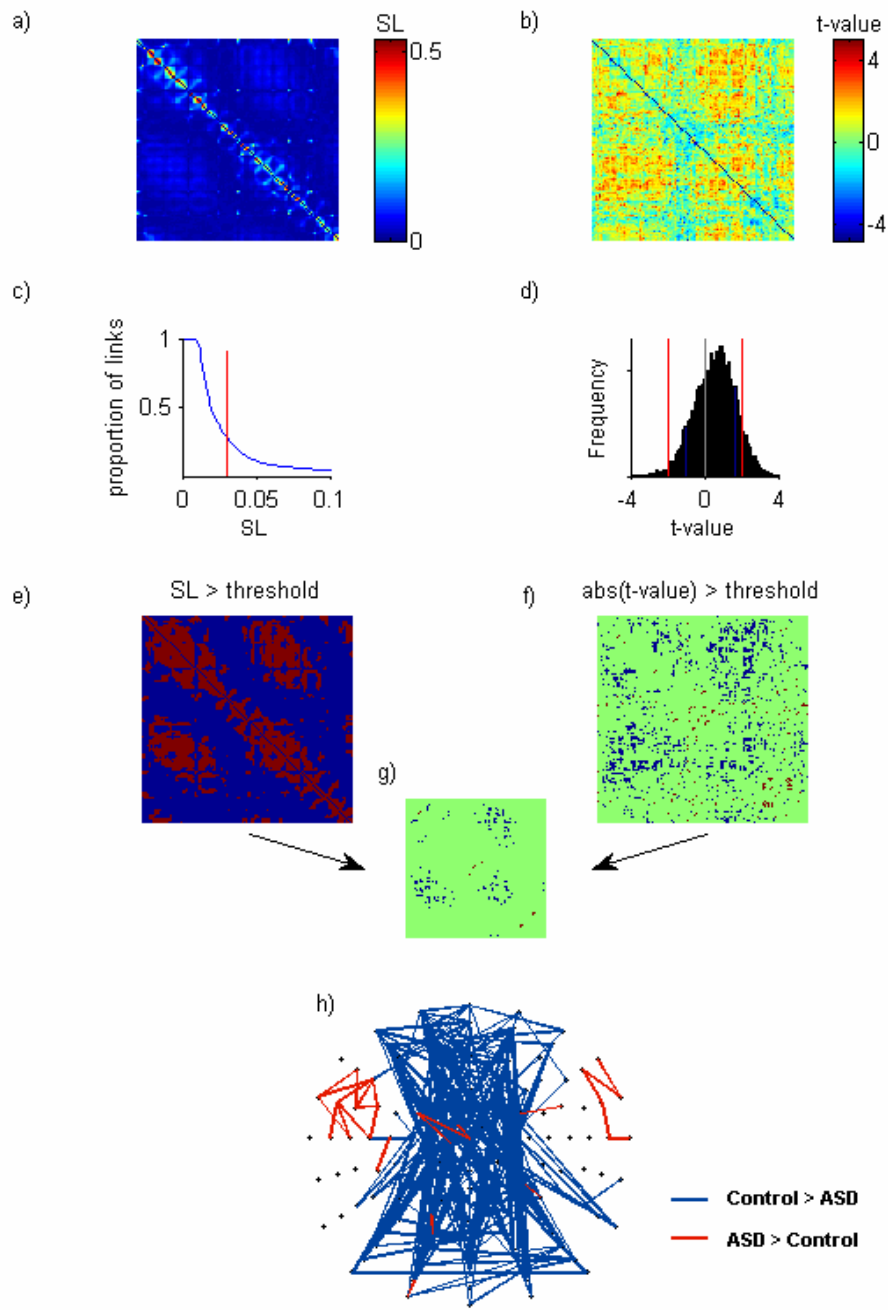

**Figure 1.** Differences in connectivity between ASD and Control groups. a) SL-matrix, averaged for all participants in this study. b) t-values of the SL difference between ASD and controls for each pair of channels. A red (blue) colour in the matrix indicates that SL is increased (decreased) in control compared to ASD population. c) Proportion of links in the SL-matrix remaining after appliance of different SL-value filters. The red line shows the threshold chosen for the analysis. d) Distribution of t-values. Red lines indicate the thresholds chosen for the analysis. e) Binary matrix showing links exceeding the SL threshold. f) Resulting matrix of thresholding the t-value matrix of fig 1b. Values are 1, 0 or -1. g) Combination of both filters gives the matrix valued in 1 and -1 for those links surpassing both filters. h) Topography of the links exceeding the threshold: Blue lines show connections significantly higher in controls, red lines show connections significantly higher in ASD.

To further quantify these observations, and to explore in a statistical manner whether fronto-occipital interactions are greater in control group and lateral-frontal interactions in the ASD group, we defined four different regions: Mid Frontal, Frontal Right Frontal Left and Occipital (Figure 2a). We then measured global connectivity across regions (including the connectivity of a region with itself), performing a t-test comparing the SL value of the weighted SL-matrix for all pairs of electrodes of the corresponding regions. This analysis revealed that local connections in Mid Frontal decrease in ASD compared to controls (t-test: $t(1,9)=3.02$; $p=0.01$), as well as the long connections between Mid Frontal and Occipital (t-test: $t(1,9)=2.21$; $p=0.05$) (Figure 2b).

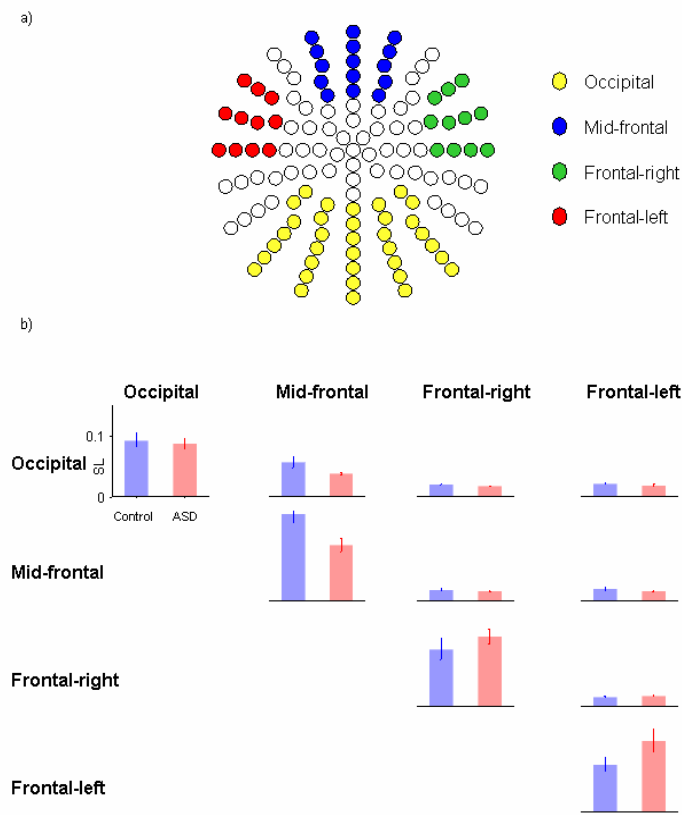

**Figure 2.** SL averaged across different regions. a) Scheme of the four regions defined for this analysis. b) SL connections for all pair or regions (connections are symmetric son only the upper-diagonal triangular matrix is shown). Comparisson within the diagonal show within region connections. Local connections in mid frontal and long connections between mid frontal and occipital are diminished in ASD compared to controls. Local connections in both Lateral Frontal areas are enhanced in ASD.

On the contrary, local connections in both Lateral Frontal areas are enhanced in ASD, being significant only in the Frontal Left (t-test: $t(1,9)=-2.35$; $p<0.05$), not in the Frontal

Right (t-test: $t(1,9)$=- 0.82; p>0.1). All other combinations of regions (Figure 2b) showed no statistical diferences.

The previous analysis showed consistent and topographically organized differences in SL between ASD and control groups, suggesting that a distinct pattern of dynamical connectivity may be related to the physiopathology of ASD. A more severe test to this hypothesis involves examining progressive changes in connectivity with a continuous progression of ASD. To examine whether the observed difference in connectivity progressed with ASD severity, we measured the correlations between ADOS score – which indexes ASD severity and varied from 7 to 16 within our population- with SL-connectivity of each pair of electrodes. Figure 3a shows two representative examples of SL pairs with a negative (blue) and a positive (red) correlation with ADOS score. The topographical distribution of the correlation of ADOS within the ASD population followed the same pattern than the difference between ASD and control groups: medial electrodes are negatively correlated with ADOS (and hence their SL is greater with decreasing levels of ASD and progressing to control population). On the contrary, lateral electrodes, predominantly for short local connections have SL values which increase with ADOS score. Note that also coherently with the group analysis; the global trend is that negative correlations (blue edges, Figure 3b) are more prominent indicating that, on average, SL connectivity increases with decreasing levels of ASD. To test quantitatively the hypothesis that short-range connections are overweighed in ASD and long-range connections are scanter, we measured the distribution of correlation coefficients for all pairs of electrodes at any given length[1] (Figure 3c). Positive correlations (SL increases with ADOS score) were very significant only within a very short range. On the contrary negative correlations (SL increases with decreasing ADOS score) were broadly distributed and extended over distant pairs of electrodes (Figure 3c). This scaling effect becomes clearer in a more quantitative manner when considering the mean value of positive and negative correlations averaged across all pairs at a fixed distance (Figure 3d). These results confirm our findings based on group analysis: ASD connectivity is overall of shorter range and dominantly localized to lateral region of the brain, with a deficit of medial occipito-frontal connections.

---

[1] Here we considered the planar distance between pairs of electrodes. This is only an approximation as interactions reflect coherent sources in a 3D volume. For the purpose of this analysis, were we merely want to explore broad scaling properties, this approximation seems adequate. Analyzing distributions considering spherical distances yielded virtually the same result.

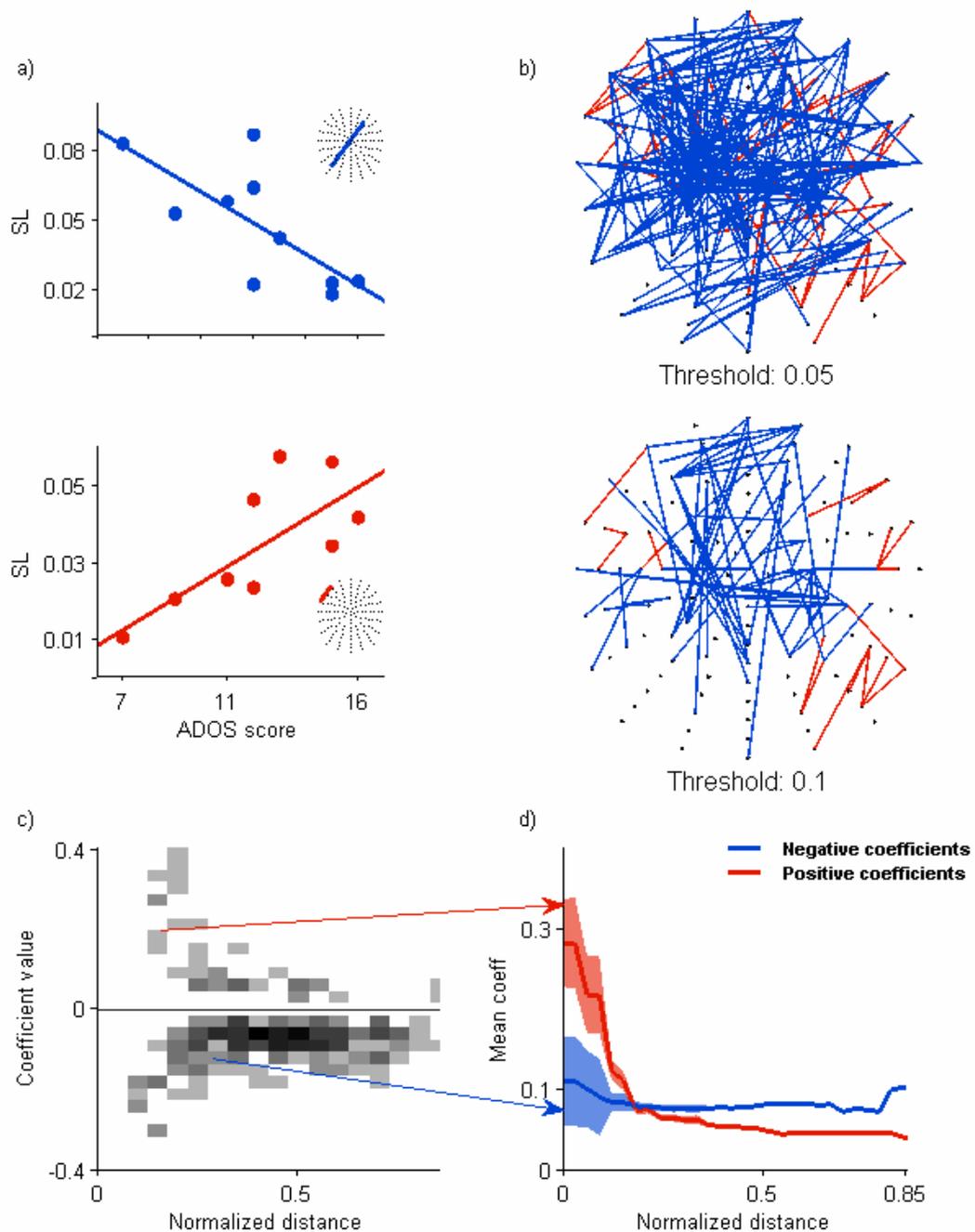

**Figure 3.** Relation between SL and ADOS. a) Representative links showing a positive and a negative relation between SL and ADOS. b) Topographic projection of the coefficients of the regression, at two different thresholds. Blue lines show negative coefficients (negative relation between SL and ASD severity) and red lines show positive coefficients (positive relation between SL and ASD severity) c) Histograms of coefficients as a function of distance between electrodes. The fraction of positive coefficients decays monotonically as the distance increases. Negative coefficients distribution is more homogeneous and remains significant at longer distances. d) Mean value of correlation coefficients as a function of distance. For short distances, correlations are on average positive, indicating that short range connections increase with ADOS. At long distances, correlations are on average negative indicating that long range connections decrease with ADOS

Our results indicate reliable differences in the connectivity pattern of ASD and control groups. We next investigated whether these differences resulted in network topologies which may have consequences in properties of information flow in the ASD and control group. Using the Kamada Kawai algorithm (Kamada & Kawai, 1989), we embedded the ASD and control networks, showing the 1000 strongest connections, in the two-dimensional plane (Figure 4a; visualizations of the networks at fixed threshold gave qualitatively the same results, see Supplementary Figure 2). By simple inspection, it is evident that the networks are qualitatively different. Control network presents a central core of nodes –composed mainly by Mid-Frontal and Occipital electrodes, considering the regions defined in Figure 2a. The ASD network is homogeneously connected, has a larger diameter and appears to be more modular and less clustered.

To quantify these observations we used four canonical graph theory metrics: Degree (K), Characteristic Path Length (L), Clustering Coefficient (C) and Modularity Index (MI). The degree (K) of the network, which constitutes its simplest statistical indicator, simply measures the average number of neighbours of each node (Figures 4b and 4c). As expected, K diminished as the threshold increases, disconnecting nodes and diminishing the size of the network. To investigate the effect of ASD on K, we submitted the K values to an ANOVA with group (control or ASD) and T (binned in 8) as independent factors. Results revealed a significant effect of group ($F(1,1)=14.48$; $p<0,01$) as well as for threshold ($F(1,7)=662$; $p<0,01$), and a significant interaction between both factors ($F(1,1)=12.15$; $p<0,01$). This shows that K was higher for the ASD group and that this effect is not invariant for all thresholds (Figure 4b). To further quantify where the differences between groups are located, we conducted a bootstrap analysis to compare K at every threshold. We found that for an intermediate range of T values (0.034-0.093), the degree was significantly higher in controls than in ASD, the most significant difference found for T = 0.056 (bootstrap test, P<0.01). At this threshold, we explored the topography of K for both groups. The scalp in Figure 4c shows the difference between scalps of both groups (Control – ASD). As with our previous findings, while the main finding is that K increases for Controls compared to ASD *on average*, it displays a rich topographical distribution with areas showing larger K and areas showing smaller K than ASD: Control group shows larger K than ASD in the Mid Frontal and Occipital areas, and smaller K in the lateral Midline, a distribution consistent with the observed pattern in figures 1 and 2 (green dots mark electrodes where $K_{control} > K_{ASD,}$; pink dots mark electrodes where $K_{ASD} > K_{control}$, $p<0.01$).

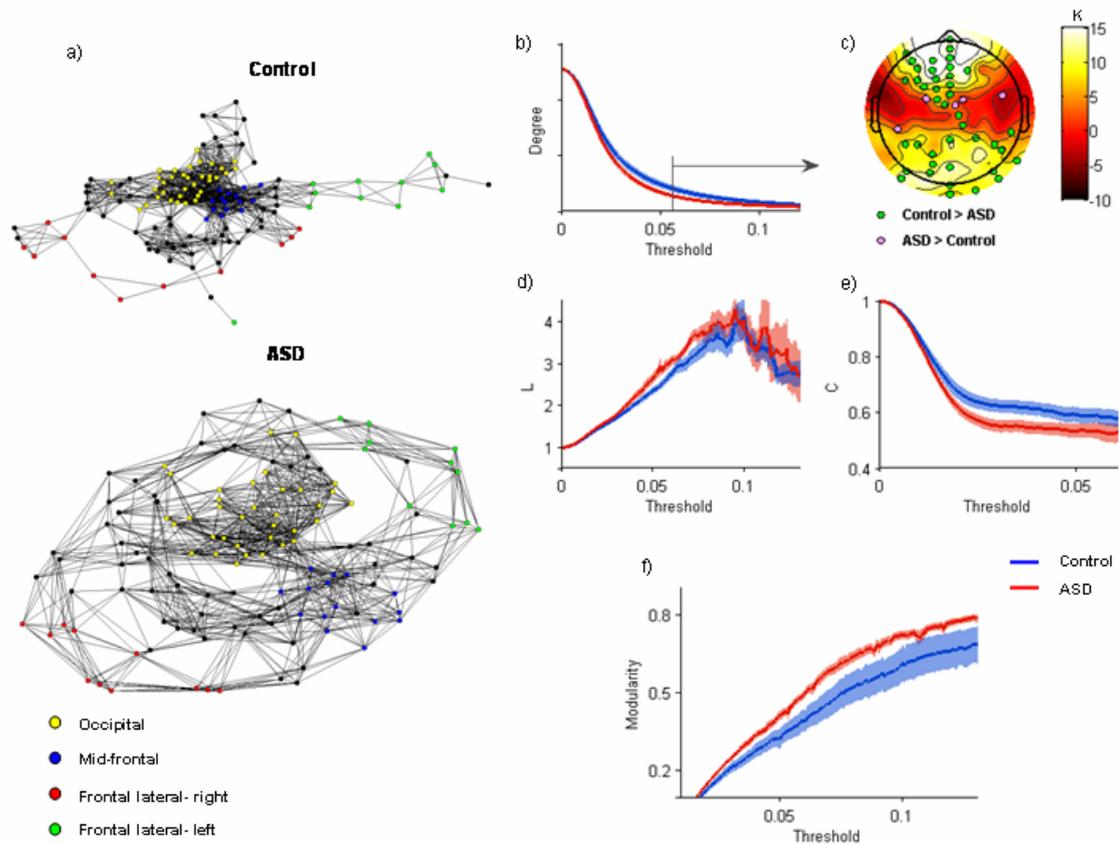

**Figure 4.** Topology. of ASD and control functional connectivity networks: a) minimal energy plots of the average networks for both groups. Colors represents the four electrode groups defined in figure 2. b) Average degree as a function of threshold. Control group shows higher degree than ASD. b) topographic map of the degree at threshold = 0.056. Green dots indicate electrodes where $K_{control} > K_{ASD}$. Pink dots indicate electrodes where $K_{ASD} > K_{control}$, P<0.01) d) Characteristic path length (L) as a function of threshold. Control group shows lower L than ASD. e) Clustering Coefficient (C) as a function of threshold. Control group shows higher C than ASD. f) Modularity as a function of threshold. ASD group shows higher modularity than Controls.

To quantify the notion of homogeneity we measured the distribution of K at different nodes (simply comparing the max and min K, a standard deviation of the distribution yielded the same results). Variations in K were less pronounced in ASD networks. At a fixed threshold the relation min(K) / max(K) for a given subject is larger in ASD than in Controls (mean controls = 0.02; mean ASD = 0.05; t(1,9): 2.39; p<0.05) demonstrating a more homogeneously connected network in ASD.

The overall dependence of the Characteristic Path Length with T also followed a well known behaviour (Figure 4d). As T increases, less edges remain and hence L increases. For very high values of T, the graph disconnects in several components, only short-range links remaining and hence L starts decreasing. To investigate the effect of ASD on Characteristic Path Length, we submitted L to an ANOVA with group (control or

ASD) and T (binned in 8) as independent factors. Results revealed a significant effect of group ($F(1,1)=8.56$; $p<0,05$) as well as for threshold ($F(1,7)=186$; $p<0,01$), and a significant interaction between both factors ($F(1,1)=4.91$; $p<0,01$), showing that L was higher for the ASD group (Figure 4d). To further quantify at which thresholds differences between groups were significant, we conducted a bootstrap analysis to compare L at every threshold. At intermediate range of values of $T$ (0.042-0.073), L is significantly larger for the ASD group, the most significant difference found at T = 0.056 (bootstrap test, $p< 0.01$).

We performed the same analysis to examine whether Clustering coefficient of ASD and Control networks differed (Figure 4e). We submitted C to an ANOVA with group (control or ASD) and T (binned in 8). Results revealed a significant effect of group ($F(1,1)=120.23$; $p<0,01$) as well as for threshold ($F(1,7)=295$; $p<0,01$), and a non significant interaction between both factors ($F(1,1)=1.88$; $p>0.05$), showing that the C was higher for the ASD group (Figure 4e). Posthoc bootstrap analysis comparing L at every threshold showed that, at intermediate range of values of $T$ (0.017-0.044), C is significantly larger for the control group; the most significant difference is found for T = 0.032 (bootstrap test, $p < 0.01$).

Much effort has been devoted to the study of statistical indicators of networks, particularly the Characteristic Path Length and the Clustering Coefficient. An ubiquitous present topological network usually referred as small-world, which has a relatively short (compared to random networks) Characteristic Path Length and high Clustering Coefficient has been shown to be optimal for information transfer and storage (Sporns & Zwi, 2004). Our combined findings (increased L and decreased C in ASD compared to control networks) indicate that the ASD network topology is consistently farther from being a small world than control network.

A direct consequence of the lack of long-range connections found in ASD is that cortical areas may become relatively isolated from each other, resulting in turn in a more modular organization (Galvao et al., 2010; Gallos, Song, Havlin, & Makse, 2007). To assess this in a quantitative manner, we estimated the Modulation Index (MI) that estimates the tendency of a network to split into modules. The dependence of MI with T follows a trend similar to L (Figure 4f): as T increases, less edges remain and MI (and the number of actual modules) increases. To investigate the effect of ASD on MI, we submitted this data to an ANOVA with group (control or ASD) and T (binned in 8) as independent factors. Results revealed a significant effect of group (F(1,1)=40.76; p<0,01) as well as for threshold (F(1,7)=220; p<0,01), and a significant interaction between both factors (F(1,1)=18.64; p<0,01), showing that the MI was higher for the ASD group (Figure 4f). To further quantify where the differences between groups are located, we conducted a bootstrap analysis to compare MI at every threshold. We found that for a very wide range of values of $T$ (0.005-0.14), MI is significantly larger for the ASD group, the most significant difference found at T = 0.065 (bootstrap test, P < 0.01)

## 4. Discussion

The main purpose of this study was to characterize and compare resting state functional brain networks in ASD and a neurotypical subjects. We studied networks derived from stationary EEG data filtered at the delta band which has been shown to covary with BOLD fluctuations in the resting state (He & Raichle, 2009; He, Snyder, Zempel, Smyth, & Raichle, 2008). We observed reliable and consistent differences in the connectivity patterns of both groups. ASD subjects showed a lack of long-range, fronto-frontal and fronto occipital connections, and an enhancement of local, lateral frontal connections. While the spatial resolution of EEG is limited, the topographical analysis reported here can only be understood as referring to broad cortical regions. With this caveat and note of caution, our results are consistent with fMRI data showing diminished or lack of diminished connectivity in the midline, more specifically between the medial prefrontal cortex and precuneus (Courchesne & Pierce, 2005; Hughes, 2007; Kana, 2006; Weng, 2010). Similarly, our observation of increased connectivity in prefronto-lateral nodes might be related with a lack of inhibition in dorsolateral Prefrontal cortex and altered connectivity of the Anterior Insula, (Di Martino et al., 2009; Kennedy, Redcay, & Courchesne, 2006; Weng, 2010).

Beyond broad group differences, our observation of a coherent topographic representation in a parametric measure of ASD determined by the ADOS score is indicative of a gradual change of network properties with increasing severity of the syndrome. Our results also revealed global trends related to proximity of correlations and ASD: for increased ASD severity, local connections increased monotonically and long range connections decreased. This global trend is also inline with fMRI results relating the severity of ASD and fMRI correlation (Di Martino et al., 2009; Kennedy, Redcay, & Courchesne, 2006; Weng, 2010).

While connectivity by itself conveys an interesting and informative measure, its main relevance lies in the implications that it has for global function (information transfer, functional specificity) of the network. Recent mathematical efforts have established connecting bridges between connectivity measures and functional properties of the emergent network (Barabasi, 2009; Galvao et al., 2010; Gallos, Song, Havlin, & Makse, 2007; Sporns & Zwi, 2004). Networks which establish an optimal balance between local specialization and global integration -displaying the highest "complexity" (Sporns & Zwi, 2004)- have characteristic small-world properties. Hence, our findings suggest that the ASD functional networks – at least as revealed by low frequency connectivity – may result in a less efficient and optimal scaffold for information processing and storage.

Cortico-cortical connections can be roughly classified in two main groups (Schroeder & Lakatos, 2009; Sporns & Zwi, 2004): local connections linking neurons in the same cortical area (critical in generating functional specificity i.e., information) and long-distance connections between neurons of different cortical regions, that ensure that distant cortical sites can interact rapidly to generate dynamical patterns of temporal correlations, allowing the integration of different sources of information into coherent behavioural and cognitive states (Bressler, 1995; Friston, 2002; Nicoll, Larkman, & Blakemore, 1993; Sporns & Zwi, 2004). The differences we found in this study suggest that this compromise is unbalanced in ASD. The reduced long range connections may provide a physiological measure for the lack of proper integration of information observed in ASD (Frith, 1989). In this sense, the organization of the whole brain networks might be related with the known differences in information processing between typical and ASD individuals.

A ubiquitous aspect of brain function is its modular organization, with a large number of processors (neurons, columns or entire areas) working in parallel. The workspace theory argues that a distributed set of neurons with long axons provides a transient global "broadcasting" system enabling communication between arbitrary and otherwise not directly connected brain processors (Baars, 1988; Baars, 2005; Dehaene & Naccache, 2001; Frith, 1989). While at this stage merely speculative and requiring further investigation, these findings suggest that ASD individuals may have an atypical workspace system, revealed in enhanced local connectivity, a more homogeneous network lacking hubs and central nodes, and a more modular organization. While the workspace system conveys brain function with a flexible communication protocol this comes at a cost: it is slow and intrinsically serial (Pashler & O'Brien, 1993; Sergent, Baillet, & Dehaene, 2005; Sigman & Dehaene, 2008; Zylberberg, Fernandez Slezak, Roelfsema, Dehaene, & Sigman, 2010). Hence, a functional manifestation of the ASD network might be to favour more parallel processing of information – being the cost of this enhancement the lack of behavioural flexibility, core symptom of individuals with ASD- an idea that resonates with the well known skills and handicaps of individuals with ASD, such as a detailed perception at the expense of a poorer integration into a big picture and, in rare cases, extraordinary performances in tasks such as the numerosity skill or calendar computation -typical of many autistic savants (Dakin & Frith, 2005; Mottron, Dawson, Soulieres, Hubert, & Burack, 2006; Thioux, Stark, Klaiman, & Schultz, 2006). If this speculations were true, ASD symptoms, its enhancements and handicaps, could be the symptoms of the lack of a proper workspace system. Future work should clarify whether the similitude observed in ASD we point here -more parallel processing/less behavioural flexibility on the one hand; more parallel low frequency EEG network/lack of a marked central core in the low frequency EEG network on the other - are indeed related.

**Acknowledgements**

MS and PB are supported by the Human Frontiers Science Program. BW is supported by the CNRS. The authors want to thank Iñaki Landerreche for providing help with the code to calculate SL**.**

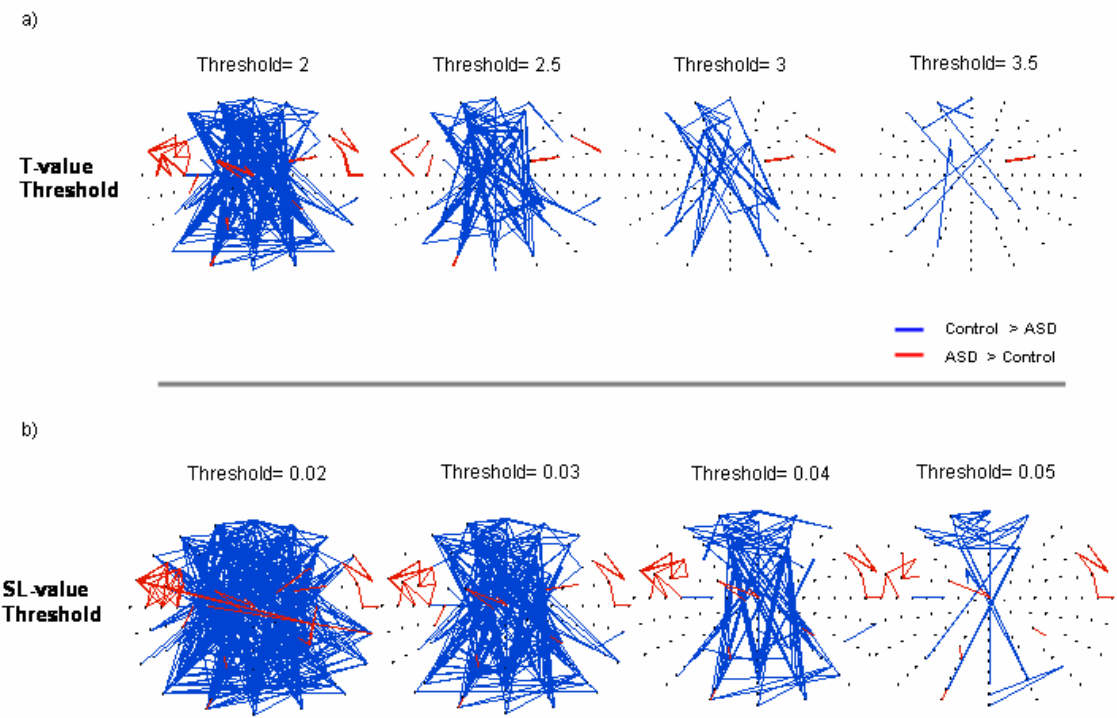

**Supplementary Figure 1.** Pairs of electrodes displaying differences in connectivity between ASD and control groups at several thresholds.. Blue lines: conections weaker in ASD than in controls. Red lines: connections stronger in ASD than in controls. a) Different values for the filter of T values (SL-value =0.03). b) Different values for the filter of SL-value (t-value =2).

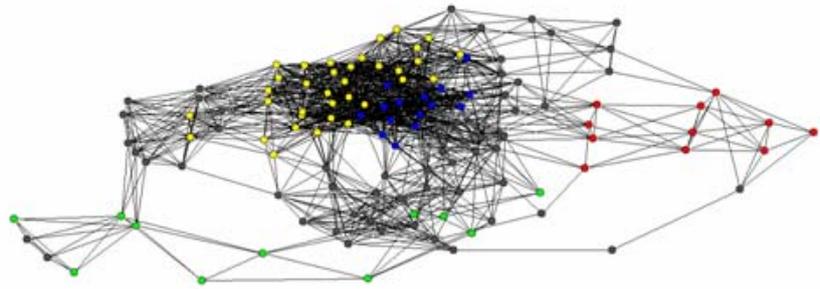
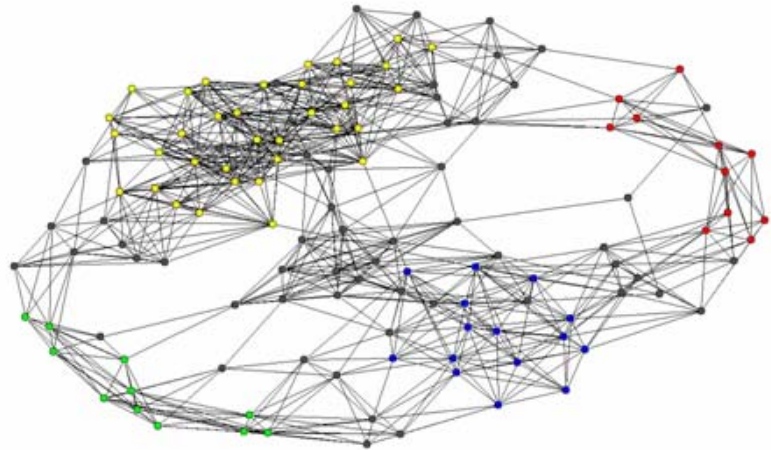

**Supplementary Figure 2.** Visualization of ASD and control networks for a fixed threshold of 0.05. The topography of the networks is qualitatively equal to the topography when the degree is fixed.